# A bias test for heteroscedastic linear least-squares regression

Eric Blankmeyer

Email eb01@txstate.edu

August 2025

A correlation between regressors and disturbances presents challenging problems in linear regression. Issues like omitted variables, measurement error and simultaneity render ordinary least squares (OLS) biased and inconsistent. In the context of heteroscedastic linear regression, this note proposes a bias test that is simple to apply. It does not reveal the size or sign of OLS bias but instead provides a statistic to assess the probable presence or absence of bias. The test is examined in simulation and in real data sets.

## Introduction

A correlation between regressors and disturbances presents challenging problems in linear regression. Issues like omitted variables, measurement error and simultaneity render ordinary least squares (OLS) biased and inconsistent (Greene 2003, 74-83, 148-149, 378-381; Basu 2020). In the context of heteroscedastic linear regression, this note proposes a bias test that is simple to apply. It does not reveal the size or sign of OLS bias but instead provides a statistic to assess the probable presence or absence of bias. The test is examined in simulation and in real data sets.

In a linear model the higher moments of the regressors and/or the residuals may enable the identification of OLS bias. For example Dagenais and Dagenais (1997) and Erickson and Whited (2002) derive instrumental variables from measures of skewness. On the other hand Lewbel (2012) and Milunovich and Yang (2018) make use of heteroscedasticity, as does this note.

OLS bias occurs when the unobservable disturbances are correlated with a regressor. But the correlation between the OLS residuals and each regressor is identically zero so a test for bias cannot be based on that moment condition. An alternative is another linear regression method for which that moment condition holds approximately but not identically under the null hypothesis of unbiasedness. Potential candidates include various robust estimators which can in principle be interpreted as weighted least squares procedures; the weights are non-linear functions of the data. This paper focuses on regression by least absolute deviations (LAD), an important instance of quantile regression (Koenker 2005, 2011; Portnoy and Walsh 1992). LAD estimates the median of a dependent variable conditional on the values of the regressors.

Consider a sample of n observations (x,y) on the bivariate linear model

$$y = \alpha + \beta x + u ,  \qquad (1)$$

where the disturbances u may or may not exhibit heteroscedasticity and may or may not be correlated with x. Let b denote the LAD estimate of β, and let r denote the Pearson correlation between x and the LAD residuals. Then Fisher's transformation (Anderson 1984, 123; Cox 2008) is

$$z = 0.5(\ln(1+r) - \ln(1-r)) , \qquad (2)$$

which has asymptotically a normal distribution with expectation zero and standard deviation $\sigma_z$ under the null hypothesis of unbiasedness—that is, the correlation between x and u is zero. The null hypothesis is rejected if

$$zstat = z/\sigma_z \qquad (3)$$

is statistically significant at conventional levels, e. g., |zstat| > 1.96.

On the assumption that x and the LAD residuals have a bivariate normal distribution, the estimate of $\sigma_z$ is $1/\sqrt{(n-3)}$. But since that assumption makes no allowance for nonspherical disturbances u, I use simulation and the bootstrap to estimate $\sigma_z$ and to produce confidence intervals for zstat.

**Three simulations**

The large-sample performance of the bias test is examined in three simulations where w ~ N(μ,σ) denotes a gaussian random variable w with expected value μ and standard deviation σ. Each simulation reports the average values of b, r and zstat when the sample of n = 500 observations is replicated five thousand times.

OLS bias due to an *omitted variable* is explored in Table 1. The linear model is

$$y = α + βx + δv + u, \qquad (4)$$

where α = 0 and β = δ = 1. Moreover x ~ N(0,2), u ~ N(0,1), and the omitted regressor v = 0.5N(0,2) + λx. If λ = 0, the OLS regression of y on x is inefficient but unbiased. There is no heteroscedasticity, b correctly estimates β, and zstat = 0.006.

**Table 1. Omitted variable simulation**
n = 500

|  |  | heteroscedasticity | |
|---|---|---|---|
|  |  | Yes | No |
| Are x and v correlated ? | Yes | b = 1.339<br>r = 0.182<br>zstat = 4.580 | b = 1.500<br>r = -0.000<br>zstat = -0.014 |
|  | No | b= 1.000<br>r = 0.000<br>zstat = 0.011 | b= 1.000<br>r = 0.000<br>zstat = 0.006 |

Heteroscedasticity is introduced when the disturbance is reformulated as $u_i \sim N(0,|v_i|)$, and the test again confirms that b estimates β accurately since zstat = 0.011. However if λ = 0.5, x and v are positively correlated, and the omission of v is expected to bias b upward. Indeed b = 1.339, and the bias is signaled since zstat = 4.580.

So in the three cases just examined, the test for bias points to the correct conclusions. However the fourth situation –no heteroscedasticity and λ = 0.5—involves a failure of identification: the upward bias in b is not reflected in zstat.

Table 2 explores OLS bias due to *measurement error*. Equation (1) is parameterized by α = 0, β = 1 and $x \sim N(0,2)$. When x is uncorrelated with the disturbance $u \sim N(0,1)$, this is the case of no measurement error and no heteroscedasticity; and b correctly estimates β. Moreover no bias is detected since zstat is -0.001. Heteroscedasticity arises when the disturbance is restated as $u_i \sim N(0,|x_i|)$. Measurement error is introduced when $ux \sim N(0,1)$ is subsequently added to x. When both measurement error and heteroscedasticity are present, b is smaller than β ("attenuation"); and zstat is statistically significant at 2.929, signaling the presence of bias. But in the case of measurement error without heteroscedasticity the downward bias in b is not reflected in zstat.

**Table 2. Measurement error simulation**
n = 500

|  |  | heteroscedasticity | |
|---|---|---|---|
|  |  | Yes | No |
| measurement error | Yes | b = 0.665<br>r = 0.138<br>zstat = 2.929 | b = 0.800<br>r = -0.001<br>zstat = -0.022 |
|  | No | b = 1.002<br>r = -0.000<br>zstat = -0.008 | b = 1.000<br>r = -0.000<br>zstat = -0.001 |

Table 3 summarizes the effects of *simultaneity bias* in a perfectly-competitive market for an agricultural commodity (Blankmeyer 2013). The log-linear demand function includes two endogenous variables, the price of the commodity and the quantity demanded; three exogenous variables –- household income, the price of a substitute commodity, and the price of a complementary commodity; and a random disturbance $u_d$. The log-linear supply function includes the price and the quantity supplied; the exogenous variables rainfall, the price of fertilizer, and the ambient temperature; and a random disturbance $u_s$.

A researcher wants to estimate the price elasticity of demand, whose "true" value is -1. The model implies that the price and $u_d$ are indeed correlated so OLS cannot estimate the price elasticity consistently (e. g., Greene 2003, 378-379). Table 3 shows that the test detects the bias when $u_d$ is heteroscedastic (zstat = 4.307) but fails to do so when $u_d$ is homoscedastic (zstat = -0.013).

<div align="center">

**Table 3. Demand elasticity simulation**
n = 500

</div>

|  |  | heteroscedasticity | |
|---|---|---|---|
|  |  | Yes | No |
| simultaneity bias |  | b = -0.699 | b = -0.481 |
|  | Yes | r = 0.148 | r = 0.000 |
|  |  | zstat = 4.307 | zstat = -0.013 |

## The administrator's salary

    A report of the Texas Health and Human Services Commission (2002) provides annual data on the administrator's salary in 842 nursing facilities operated for profit. In a log-linear model the salary is regressed on variables that affect the facility's profitability and presumably the manager's compensation: occupancy rate, revenue, area (in square feet) and staff size. For OLS the regressors are statistically significant, and heteroscedasticity is confirmed by the Breusch-Pagan test (e.g., Greene 2003, 223-225). No regressor is significantly correlated with the LAD residuals so bias is not detected (Table 4).
    However if the occupancy rate is dropped from the model, zstat = 3.575 when the LAD residuals are correlated with revenue, reflecting the omitted-variable bias. Furthermore the bootstrap confidence interval indicates that the probability of observing an insignificant zstat when OLS is in fact biased would be less than 5 percent.

**Table 4. The administrator's salary**
(the dependent variable is ln salary,
standard errors are under coefficients*)
n = 842

|  | OLS | OLS |
|---|---|---|
| occupancy rate | 0.303 | omitted |
|  | 0.075 | variable |
|  |  |  |
| ln revenue | 0.421 | 0.664 |
|  | 0.047 | 0.075 |
|  |  |  |
| ln area | -0.086 | -0.263 |
|  | 0.036 | 0.047 |
|  |  |  |
| ln staff size | -0.095 | -0.099 |
|  | 0.035 | 0.036 |
|  |  |  |
| zstat for: |  |  |
| occupancy | 0.378 | --- |
| ln revenue | 1.289 | **3.575** |
| ln area | 1.315 | 1.433 |
| ln staff size | 0.531 | 1.233 |

\* The standard errors for coefficients are hetero-
scedasticity- and autocorrelation-consistent (HAC),
Newey-West version.
The zstats are bootstrap estimates.

## Household expenditures on food

The data set "VietnamH" (Croissant 2015) is a 1997 survey of expenditures by 5,999 Vietnamese households. Outlays for food can be modeled as a function of total expenditures, household size and other factors. OLS might be biased since total expenditure "and its components…are endogenous to the consumer and are determined simultaneously" (Liviatan 1961, 336). Liviatan argues that OLS will be skewed downward when the dependent variable is a relatively stable component of expenditure like food while an upward bias should be expected for highly variable items such as major appliances.

The Breusch-Pagan test strongly confirms heteroscedasticity. In Table 5 the OLS elasticity of food outlays with respect to total household expenditures is 0.659, but it is probably biased since zstat = -3.667 when the LAD residuals are correlated with total expenditures.

Table 5. Food expenditure elasticity
(the dependent variable is ln food expenditure,
standard errors are under coefficients*)
**n = 5999**

|                      | OLS    |
|----------------------|--------|
| ln total expenditure | 0.659  |
|                      | 0.012  |
| household size       | 0.043  |
|                      | 0.003  |
| gender (male = 1)    | 0.056  |
|                      | 0.009  |
| farm (yes = 1)       | 0.037  |
|                      | 0.011  |
| zstat                | **-3.667** |

* The standard errors for coefficients are are heteroscedasticity- and autocorrelation-consistent (Newey-West version)

The zstat is a bootstrap estimate.

**The demand for nursing services**

       Drawing on a data base of the Texas Health and Human Services Commission (2002), I estimate the demand curve for nursing services in Texas long-term care facilities. The sample is comprised of 824 for-profit nursing homes licensed by the state in 2002. According to the textbook model of a competitive market, the demand for a resource depends on its price, on the usage levels of other inputs, and on the price of the good or service to be produced–in this case a nursing facility's average revenue per resident day. In conjunction with the supply curve for the resource, this resource-demand function determines the wage rate.

       I focus on the demand curve for the services of licensed vocational nurses (LVN), also called licensed practical nurses, who have typically completed one or two years of formal training and who work under the supervision of registered nurses (RN) and physicians. In the log-linear model the jointly endogenous variables are the total LVN hours worked during 2002 and the average hourly LVN wage rate in each facility. The included exogenous variables are the total hours worked by RN, by nurse's aides (AIDE), and by laundry and housekeeping personnel (L+H) together with the number of beds in the facility and the revenue per resident day.

       The Breusch-Pagan test confirms heteroscedasticity. In Table 6 each OLS coefficient is statistically significant at conventional levels except for RN hours. The demand is inelastic, -0.396, but zstat = -2.387. The endogeneity of hours worked and the hourly wage could produce simultaneity bias. Excluded exogenous variables would presumably be the determinants of the LVN supply curve, e. g., the LVN's age, the number of young children in the family, a spouse's income, and the local cost of living. However, these potential instrumental variables are unavailable. Blankmeyer (2022) uses canonical correlation to estimate the LVN demand elasticity at -0.649.

**Table 6. The LVN demand model**
(the dependent variable is ln LVN hours,
standard errors are under coefficients*)
n = 824

|  | OLS |
|---|---|
| ln LVN hourly wage | -0.396 |
|  | 0.104 |
|  |  |
| ln number of beds | 0.158 |
|  | 0.040 |
|  |  |
| ln RN hours | 0.045 |
|  | 0.033 |
|  |  |
| ln aide hours | 0.669 |
|  | 0.072 |
|  |  |
| ln L+H hours | 0.138 |
|  | 0.058 |
|  |  |
| ln revenue per resident-day | 0.350 |
|  | 0.075 |
|  |  |
| zstat | **-2.387** |

* HAC standard errors with Newey-West /Bartlett window
  are reported for the regression coefficients, and zstat
  is a bootstrap estimate.

## An earnings equation

An earnings equation explains workers' wages as a function of their schooling, job experience, ethnicity, location, and other factors (e. g., Heckman et al. 2003). The data set CPS88 (Kleiber and Zeileis, 2015) draws on a U. S. Census survey of 28,155 male workers who were not self-employed. The Breusch-Pagan test confirms heteroscedasticity, and the OLS regression is displayed in Table 7. In particular, the rate of return to an additional year of education, 9.3 percent, is broadly consistent with the findings of many other studies.

However a perennial concern is the omission of a regressor to control for a worker's skills and innate ability, which are difficult to quantify. Presumably the regressor would be positively correlated with both education and earnings, so its omission would skew the OLS coefficient for education upwards. Indeed the zstat in Table 7 is -5.600, strongly indicative of bias.

Table 7. The earnings equation
(the dependent variable is ln weekly wage, standard errors are under coefficients)
n = 28,155

| | OLS |
|---|---|
| education (years) | 0.093 |
| | 0.001 |
| experience (years) | 0.017 |
| | 0.000 |
| ethnicity | 0.218 |
| (T = caucasian) | 0.012 |
| smsa | 0.157 |
| (T = yes) | 0.008 |
| Northeast region | 0.038 |
| | 0.010 |
| South region | -0.050 |
| | 0.009 |
| West region | 0.018 |
| | 0.010 |
| part time | -1.071 |
| (T = yes) | 0.012 |
| zstat (bootstrap) | **-5.600** |

## Income and infant mortality

The dataset "UN" (Fox and Weisberg 2015) reports infant mortality (deaths per 1,000 live births) and per-capita gross domestic product (gdp in thousand U. S. dollars) for 193 nations in 1998. The OLS regression shows that mortality decreases as income rises: the slope is -2.211 with a standard error of 0.228. Moreover heteroscedasticity is confirmed by the Breusch-Pagan test. Bias is highly probable since zstat = -4.200. Furthermore the bootstrap confidence interval indicates that the probability of observing an insignificant zstat when OLS is in fact biased would be less than 5 percent.

Does the OLS bias reflect attenuation due to *measurement error* ? It seems likely that gdp is significantly mismeasured for countries with large underground sectors, poorly-funded data collection programs and unrealistic exchange rates vis-a-vis the dollar.

## Summary and outlook

This paper offers preliminary evidence that the test can signal the presence or absence of OLS bias in heteroscedastic linear regression. Application of the bias test involves negligible marginal costs of data acquisition and computation. Indeed if the Breusch-Pagan test confirms heteroscedasticity and if zstat is not statistically significant, a search for valid instrumental variables may be avoided.

The reliability of the bias test obviously depends on the quality of the regression model. For instance, if the regressor v in equation (4) accounts for a large part of the variance in y, then the omission of v tends to degrade the model's overall goodness of fit, which may render zstat statistically insignificant despite the bias in the estimate of $\beta$. In equation (4) $\beta = \delta = 1$; but if instead $\delta = 5$, then the simulation produces zstat = 1.029 because the bootstrap standard error of z = 0.035 although $z \approx r = 0.036$.

Research is underway to examine alternatives to LAD. Simulation tentatively indicates that high-breakdown regression (Hubert et al. 2010, 2015; Maronna et al. 2006, chapter 5; Rousseeuw and Van Driessen 2006) may detect bias more effectively than LAD in challenging situations where heteroscedasticity is weak, or the sample is not very large, or the bias is not very severe.

In the examination of five real data sets I have attributed each large |zstat| to an omitted variable, simultaneity bias, or measurement error. Of course those attributions cannot be categorical since additional specification problems or data issues may also be skewing the OLS estimates.